\documentclass[acus]{JAC2000}

\usepackage{graphicx}

\begin{document}
\title{\flushright{W09}\\[15pt] 
\centering BARYON FORM FACTORS IN QCD\thanks
{Work supported by the NSF (USA) under Grant No.\ PHY-9900657.}
}

\author{Carl E. Carlson\\ Nuclear and Particle Theory Group,
Physics Department\\ College of William and Mary, Williamsburg, VA 23185,
USA}

\maketitle

\begin{abstract}
This is a summary of perturbative QCD calculations of baryon
form factors.  For $e^+ e^-$ going to baryon-antibaryon pairs, normalized
calculations are available and reported for the entire ground state
octet and decuplet, including off-diagonal form factors, and for the
$S_{11}$(1535)-$\overline S_{11}$.  (The latter results are new for this
report.)  We also include some explanation of how the results come to be.
\end{abstract}

\section{INTRODUCTION}

The form factors of baryons have long been studied in electron scattering,
and now we are here to discuss new opportunities that can  come from
an $e^+ e^-$ collider at moderate energies where exclusive cross sections
may be measured.  One can foresee measurements of the form factors of
baryons in the timelike region.  The richness of these measurements
should be clear.  We can in principle measure the elastic and
transition form factors of any baryon, since the baryon-antibaryon pair is
produced in the reaction.  We are not limited to baryons which exist
in stable targets.

This talk will focus on results obtained using perturbative QCD (pQCD). 
They will therefore be valid at high $Q^2$, but should at a minimum serve
as an estimate of which form factors will be big and which will be small. 
It should be emphasized that pQCD is not a model.  It is an outcome of
the real theory of the strong interactions.  The scaling  laws can be
quoted and demonstrated without approximation, and normalized calculations
proceed directly if the lowest Fock state wave functions of the quarks
inside the hadrons are known.  Some modeling is, however, needed for the
latter.  The models are not pure invention, since significant information
about (moments of, at least) the wave functions can be gotten from QCD
sum rule calculations and, less extensively, from lattice gauge theory.

We will tabulate and discuss the results for the form factors, elastic
and non-elastic, for the entire ground state baryon octet and decuplet in
the third section.  The results are taken from many places in known
literature. We will also give some new results for form factors
involving the $S_{11}$(1535), where it happens that the ingredients have
been available for some time, but had been put together for nucleon
to $S_{11}$ transitions but not for the elastic case.

The second section will contain a brief summary of how the results come
to be, including some necessary special attention to the peculiar case of
the nucleon to $\Delta$(1232) transition form factor.

\section{HOW TO DO IT}

\subsection{Basics}

At high $Q^2$, the magnetic form factor $G_M$ (but not the electric
one $G_E$) can be calculated using perturbative QCD.  The basic result is
that the form factor factors into terms representing the wave function
or distribution amplitude of the baryons and a hard scattering kernel. 
Schematically,
\begin{equation}
G_M = \int [dx][dy] \Phi(y,Q^2) T_H(x,y,Q^2) \Phi(x,Q^2)  \ .
\end{equation}

In more detail, the lowest Fock component of a proton can be represented
as~\cite{lb}
\begin{eqnarray}
|p, \uparrow \rangle &=& \int {[dx\, d^2k_T] \over (x_1 x_2 x_3)^{1/2}}
             \\ \nonumber &\times&
    \Big\{ {\psi_S(x_i,k_{iT})  \over \sqrt{6}}
         \left(2udu -uud-duu \right)_{\uparrow \downarrow \uparrow}
             \\ \nonumber &+&
   \quad    {\psi_A(x_i,k_{iT})  \over \sqrt{2}}
         \left(uud-duu \right)_{\uparrow \downarrow \uparrow}  \Big\} \ .
\end{eqnarray}

\noindent  There are two lowest Fock wave functions for isospin 1/2
particles like the nucleons, $\psi_{S,A}$. If
$k_T$ is limited and if
$Q >> \langle k_T \rangle$, one can get the factored $G_M$, using
\begin{equation}
\phi_{S,A}(x,Q^2) = \int^{\sim Q} [d^2k_T] \psi_{S,A}(x,k_T)
\end{equation}

\noindent where one expects the $Q$ dependence to be quite weak,
and results like
\begin{eqnarray}
G_M &=&  {1\over Q^4} 
\left( 16\pi \alpha_s \over 3 \right)^2 \int [dx \,dy]
    \times   \Big\{
2 T_1 \phi_S (x) \phi_S(y) 
       \nonumber \\
&+& {2\over 3} (T_1 + 2 T_2) \phi_A (x) \phi_A(y)
       + {\rm  cross\ terms} \Big\}  .
\end{eqnarray}

\noindent Here, the $1/Q^4$ and $T_{1,2}$ have come from evaluating the
hard scattering kernel, and~\cite{lb,cz80}
\begin{eqnarray}
T_2 = {1\over x_1 x_3 (1-x_1)y_1 y_3 (1-y_3)} \ ,
\end{eqnarray} 

\noindent and similarly for $T_1$. 

\subsection{Scaling and spin rules}

\begin{figure}[h]
\centering
\includegraphics[width=2.5in]{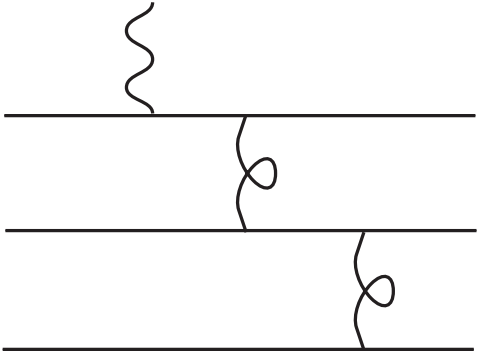}
\caption{Example of lowest order perturbation theory diagram hard
scattering kernel in baryon form factor calculation.}
\label{one}
\end{figure}

This section could be subtitled ``Properties of the hard scattering
kernel.''  There are scaling and spin selection rules that follow simply
from a few precepts.  A typical diagram for $T_H$ is shown in
Fig.~\ref{one}.

To get the power of $Q$ associated with this diagram, the rules are
\begin{itemize}
\item  $1/Q^2$ for each gluon propagator,
\item  $1/Q$ for each quark propagator,
\item  Q for each quark line.
\end{itemize}

\noindent The spin selection rules are (in part),

\begin{itemize}
\item quark helicity is conserved,
\item if a single gluon or photon is connected to a quark line, it must
have transverse polarization,
\item if two vector bosons are attached to a quark line, one must have
transverse polarization, the other must have longitudinal polarization,
\item the helicities of two quarks connected by a transverse gluon must
be opposite.
\end{itemize}

\noindent There are also definitional $Q$'s. The above rules are for
Feynman diagrams, and a form factor may be defined as a certain factor
times a basic matrix element.

Any of the rules can be violated, but it ``costs'' factors of $O(m/Q)$ or
of $O(\langle k_T \rangle/Q)$

\subsection{Application to nucleon}

For the nucleon, in the Breit frame one can show
\begin{eqnarray}
G_M &\equiv& {1\over Q \sqrt{2}} \langle p,\uparrow | 
            \varepsilon_T \cdot J | p,\uparrow \rangle
   \nonumber \\
    &\propto& {1\over Q}  \times {1\over Q^3}  = {1 \over Q^4}
\end{eqnarray}

\noindent where $\varepsilon$ is a photon polarization vector, $J$ is the
electromagnetic current operator, and the arrows are for helicity in the
Breit frame.  Further,
\begin{eqnarray}
G_E \equiv && {1\over 2 m_N} \langle p,\downarrow | 
            \varepsilon_T \cdot J | p,\uparrow \rangle
   \nonumber \\
    \propto && {\langle k_T \rangle\over Q}  \times {1\over Q^3}  
  = {1 \over Q^4}
\end{eqnarray}

\noindent For $G_M$, an extra $1/Q$ comes from its definition, but for
$G_E$ the extra $1/Q$ is due to violating a spin selection rule.  The
predicted falloff famously works well for $G_M$, beginning at $Q^2$ below
10 GeV$^2$. 


The results can also be converted into Dirac and Pauli form factors, 
\begin{eqnarray}
F_1 &=& {\tau G_M + G_E \over \tau +1} \nonumber \\
F_2 &=& {G_M - G_E \over \tau + 1} \propto {1/Q^6}, {\rm at\ high\ } Q^2,
\end{eqnarray}

\noindent where $\tau = Q^2/4 m_N^2$.  Note that $G_M$ and $F_1$ become
identical at high $Q^2$.  The prediction for $F_2$ is not yet working
well~\cite{jones}.

\subsection{Other Resonances}

This section will be mainly about notation, with numerical results coming
later.  Electromagnetic nucleon to resonance transitions are often, as in
the tables of the Particle Data Group, given in terms of scaled helicity
amplitudes---$A_{1/2}$ and $A_{3/2}$ for photons with transverse
polarization.  For convenience in writing cross section formulas, these
amplitude have been divided by the momentum of the photon causing the
transition, in the real photon limit.  This momentum is zero for an
elastic form factor, so the the $A_{1/2,3/2}$ notation is useless for
elastic transitions.  I feel it is better to always use an unscaled
helicity form factor.  With a mass factor inserted to make the form
factor dimensionless, one has~\cite{c86}
\begin{eqnarray}
2 m_N G_+ = \langle B', +{1\over 2} | \varepsilon_T \cdot J 
          | B, +{1\over 2} \rangle
\end{eqnarray}

\noindent where the $+1/2$ is in both cases a helicity in the Breit
frame. (Equivalent is the $f_+$ from~\cite{bw}, with 
$2 m_N G_+ = \sqrt{2 m_B' 2 m_B} f_+$.)  Using $G_+$ one can directly
compare elastic and off-diagonal form factors.  Useful connections are
\begin{eqnarray}
G_M = {m_N \sqrt{2} \over Q} G_+(N \rightarrow N)  \nonumber \\
A_{1/2} = e \sqrt{m_N \over m_R^2 - m_N^2} G_+ (N \rightarrow R) .
\end{eqnarray}

\noindent (where $e$ is the electric charge).  It is also possible to
define $F_1$ in a straightforward manner, asymptotically as $(m_N
\sqrt{2}/Q) G_+$, for off-diagonal transitions and use it to make
interbaryon comparisons.  

Stoler has presented plots testing the pQCD scaling for nucleon to
resonance transitions~\cite{stoler}. 
The scaling predicted by perturbative QCD works for three cases out of
four, starting safely below 10 GeV$^2$ and continuing until the data runs
out just past 20 GeV$^2$.  The exception is the 
$N \rightarrow \Delta$(1232) transition, which will get a dedicated
discussion in the next section.

%

\subsection{The Distribution Amplitudes}

This will be just a brief description of how one uses QCD sum rules to
get distribution amplitudes for resonances.  A full description can be
gotten from the original literature~\cite{cz,ks}, or from lectures
written up by the present author~\cite{drontenseattle} (who was among
those who extended the method to the Delta resonance~\cite{cp}).

The idea is to start with some function, such as,
\begin{equation}
I(q) = i \int d^4y e^{-iqy} \ \langle 0 | T O_1(y)\ O_2(0) | 0 \rangle
\end{equation}

\noindent that one can evaluate in two different ways. Before doing any
evaluation, however, one chooses at least one  of the operators to ensure
that only intermediate states of the desired quantum numbers (e.g.,
isospin 3/2, positive parity, if one is interested in the Delta) can
enter. 

One
of the ways to evaluate is a quark/gluon evaluation, which will 
depend on already fitted parameters like the density of quark pairs and
gluons in the physical vacuum, but which can be evaluated from start to
finish.  The other way is a purely hadronic evaluation, done by inserting
complete sets of intermediate states between the two operators, which
depends on the wave functions of the quarks inside the hadron.  One
actually gets moments of the distribution amplitude (the distribution
amplitude multiplied by powers of momentum fractions, and integrated),
the moment depending on the details of the operator chosen.  Then one
matches the two results to get the numerical value of the moment.  

Having all the moments is equivalent to having the wave function. 
Unfortunately, uncertainties in the evaluations build up for higher
moments, so that one only gets a few low moments.  The information is
still valuable, and allows reasonable and normalized choices for
distribution amplitudes of the various particles.  Uncertainties also
build up for any state but the lowest in a given category.  For
example, in the non-strange sector, results are only available for the
nucleon, the Delta(1232), and the $S_{11}$(1535).   The results
for the moments generally show an asymmetric distribution amplitude for
the octet baryons, wherein quarks with helicity paralleling the parent
baryon generally carry a larger that equipartition share of the
momentum.  The distribution amplitudes for the decuplet is, on the other
hand, rather symmetrical with momenta on the average evenly divided.

 
\section{NUMERICAL PREDICTIONS}

\subsection{Nucleon Form Factors}

We can be brief.  For the nucleon form factors, in the spacelike region,
the data has long been known.  Hence the data is well fit by the
theory---or else we would never of heard of the theory. 

There are criticisms of the use of pQCD at current experimental momentum
transfers.  We will hardly discuss these here.  They are based on
excising contributions where internal four-momenta squared are low, and
seeing what remains.  In my opinion, the cutoffs used are quite
pessimistic, and further do not consider that internal lines with
low four-momentum squared can be quite perturbative if they are short
range in coordinate space.  Discussion on the positive side can be found
in~\cite{plus}, and on the negative side in~\cite{minus}.

\subsection{The Nucleon $\rightarrow$ Delta}

The pQCD scaling is not seen for the $N \rightarrow \Delta$(1232)
electromagnetic transition.  Instead we have the DDR---the Disappearing
Delta Resonance, and the resonance peak sinks into the background with
increasing $Q^2$.

But we now know the distribution amplitudes $\phi_\Delta$ and
$\phi_{S,A}$ for the nucleon, and~\cite{cgs} 
\begin{eqnarray}
m_N Q^3 G_+(N\rightarrow \Delta) = \left(16\pi \alpha_s^2/9 \right)^2
          \times  \qquad \qquad    \nonumber \\    \times
     \left\{ {2\over 3} \langle \phi_A | T_1 - T_2 | \phi_S \rangle
    +  {1\over \sqrt{3}} \langle \phi_A | T_1  | \phi_A \rangle
     \right\}
\end{eqnarray}

\noindent There is a substantial cancellation between the two terms
above.  The numbers are,
\begin{equation}  \label{asymp} 
Q^3 G_+(N \rightarrow \Delta) = \left\{
\begin{array}{cl} 
0.05 {\rm GeV}^{3}  &  \quad {\rm CZ,CP} \\ 
0.08 {\rm GeV}^{3}  &  \quad {\rm KS,CP} \\
0.02 {\rm GeV}^{3}  &  \quad {\rm COZ,FZOZ \,,}
                                            \end{array}
                                            \right.  
\end{equation}

\noindent where the letter codes refer to distribution amplitudes
for the nucleon and Delta, in that order, from papers
in~\cite{cz,ks,cp,coz,fzoz,bp}.

The numbers are small.  For comparison,
\begin{equation} 
Q^3 G_+(p \rightarrow p) = {1\over m_N \sqrt{2}} Q^4 G_M 
           \approx 0.75 {\rm \ GeV}^3  ,
\end{equation}

\noindent and
\begin{equation} 
Q^3 G_+(p \rightarrow N^*(1535)) = \left\{
\begin{array}{cl} 
0.46 {\rm GeV}^{3}  &  \quad {\rm CZ,CP} \\ 
0.58 {\rm GeV}^{3}  &  \quad {\rm KS,CP \,.}
\end{array}
\right.
\end{equation}

We conclude that, in distinction to the situation for other form factors,
the leading order $N \rightarrow \Delta$ form factors are not
currently above background.  There is a corollary, which is that the spin
prediction $E_{1+} = M_{1+}$ for $Q \rightarrow \infty$ is no seen
because the leading order amplitude is not  yet dominant.  We expect, or
hope, that $Q^2 > 10 GeV^2$ will show some noticeable leading order
amplitude, with the ratio $E_{1+}/M_{1+}$ rising. There is further
discussion in~\cite{cm98,cj}.

\subsection{Other Baryons}

We come finally to present the results for a whole array of baryons. 
These predictions depend on distribution amplitudes obtained from QCD
sum rules in~\cite{cp,bp,fzoz,coz}, with special credit gong to the last
listed for having done the largest number of baryon resonances.

We quote in all cases values for $Q^4 F_1$ in GeV$^4$, calculated with a
fixed $\alpha_s = 0.3$.  First in Table~\ref{octet}, we give $Q^4 F_1$
for diagonal transitions (e.g., $\Sigma^+$ for the timelike region means
$\Sigma^+ \overline{\Sigma^+}$ = $\Sigma^+ \overline\Sigma^-$) for the
ground state baryon octet.  The prediction for the $\Sigma^0 \rightarrow
\Lambda^0$ electromagnetictransition is also given.  (For full set of
octet baryons, there are also predictions from the diquark
model~\cite{jakob}.)

\begin{table} [h]    
\begin{center}
\begin{tabular}{cc}
octet baryon & $Q^4 F_1$ GeV$^4$ \\ \hline\hline
n            & $-$0.5   \\
p            & 1.0    \\
$\Sigma^-$   & $-$0.65  \\
$\Sigma^0$   & 0.27   \\
$\Sigma^+$   & 1.19   \\
$\Lambda^0$  & $-$0.23  \\
$\Xi^-$      & $-$0.60  \\ 
$\Xi^0$      & $-$0.52  \\
$\Sigma^0 \rightarrow \Lambda^0$  &  0.54  \\
\hline
\end{tabular}
\caption{Asymptotic form factors $F_1$ for the octet baryons.}
\label{octet}
\end{center}
\end{table}

Table~\ref{decuplet} is the same but for the ground state decuplet
baryons.  These tend to be smaller than the octet because the
smoothness of the wave function gives less  strength near the end points,
where the bulk of the contributions arise.

\begin{table} [h]    
\begin{center}
\begin{tabular}{cc}
decuplet baryon & $Q^4 F_1$ GeV$^4$ \\ \hline\hline
$\Omega$        & $-$0.02   \\
$\Xi^{*-}$      & $-$0.083  \\ 
$\Xi^{*0}$      & 0.014   \\
$\Sigma^{*-}$   & $-$0.031  \\
$\Sigma^{*0}$   & 0.016   \\
$\Sigma^{*+}$   & 0.062   \\
$\Delta^-$      & $-$0.085  \\
$\Delta^0$      &  0      \\
$\Delta^+$      &  0.085  \\
$\Delta^{++}$   &  0.17   \\
\hline
\end{tabular}
\caption{Asymptotic form factors $F_1$ for the decuplet baryons.}
\label{decuplet}
\end{center}
\end{table}

And then in table~\ref{offdiagonal} we have $Q^4 F_1$ for the
ground state decuplet to octet baryon transitions, or associated
baryon-antibaryon production in the timelike region.

\begin{table} [h]    
\begin{center}
\begin{tabular}{cc}
octet-decuplet transition & $Q^4 F_1$ GeV$^4$ \\ \hline\hline
n $\rightarrow \Delta^0$             & $-$0.08    \\
p $\rightarrow \Delta^+$             &  0.08    \\ 
$\Sigma^- \rightarrow \Sigma^{*-}$   &  0.016   \\
$\Sigma^0 \rightarrow \Sigma^{*0}$   &  0.024   \\
$\Sigma^+ \rightarrow \Sigma^{*+}$   &  0.033   \\
$\Lambda^0 \rightarrow \Sigma^{*0}$  &  0.015   \\
$\Xi^- \rightarrow \Xi^{*-}$         &  0.013   \\
$\Xi^0 \rightarrow \Xi^{*0}$         &  0.024   \\
\hline
\end{tabular}
\caption{Asymptotic form factor $F_1$ for the octet to decuplet baryon
transitions.}
\label{offdiagonal}
\end{center}
\end{table}

Finally, we have table~\ref{s11} which gives the asymptotic $Q^4 F_1$
for transitions involving the negative parity $S_{11}$(1535) resonance.
A surprise is the large size of the neutral $S_{11}$ elastic form factor.

\begin{table} [h]    
\begin{center}
\begin{tabular}{cc}
transition     & $Q^4 F_1$ GeV$^4$ \\ \hline\hline
n $\rightarrow S_{11}^0$             &  0.35   \\
p $\rightarrow S_{11}^+$             &  0.7    \\ 
$S_{11}^0 \rightarrow S_{11}^0$      &  1.6    \\
$S_{11}^+ \rightarrow S_{11}^+$      &  0.17   \\
\hline
\end{tabular}
\caption{Asymptotic form factors $F_1$ for 
transitions involving the $S_{11}$(1535).}
\label{s11}
\end{center}
\end{table}

\subsection{Comments on $e^+e^- \rightarrow B \bar B$}

As is too commonly said, the relation between the spacelike and timelike
region at finite $Q^2$ requires more thought.  The data~\cite{seth} on
$F_1$ or $G_M$ for $p \bar p$ is about twice as large at timelike $Q^2$
than at the corresponding spacelike $Q^2$ in the 10 GeV$^2$ region.  A
rough estimate of the effects of quark mass and transverse momenta shows
that a factor of 2 in this $Q^2$ region in reasonable from the
theoretical side. (The estimate is along the lines of estimates in the
spacelike region showing that the mass term in the dipole parameterization
of about 0.71 GeV$^2$ is reasonable~\cite{ancientbc}.)    One can take
several attitudes to this statement.  One is that pQCD with expected
amendations does well.  Another is that is shows that we are not at the
point where masses are neglectable.  These are higher twist effects, and
there are others, including contributions of higher Fock states to the
form factors.

\section{LAST THOUGHTS}

Baryon form factors are calculable at high $Q^2$ using perturbative QCD
in either the spacelike or timelike region.  The results are
experimentally good for the nucleon elastic and nucleon to $S_{11}$ form
factors at $Q^2$ above a few GeV$^2$ spacelike, and one can explain the
lack of currently observed scaling of the nucleon to $\Delta$(1232) 
transition.  

The proposed $e^+ e^-$ machine is a vehicle to measure timelike form
factors for a host of unstable baryons.  Predictions exist to shoot at
for octet elastic, decuplet elastic, octet-decuplet off-diagonal, and
$S_{11}$ form factors.  There are also non-pQCD predictions, for example
those reported by Dubnicka {\it et al.} at this meeting~\cite{dub}, and
prediction from the diquark model by Jakob {\it et al}~\cite{jakob}.

The results will cast light on the quarkic wave functions of baryons. 
The predictions quoted here are specific to the QCD sum rule wave
functions we have used.  The actual wave functions may be different and
the form factor measurements, with the possibility of combining them with
$J/\psi, \psi, \ldots \rightarrow B \bar B$ and 
$\gamma  \gamma \rightarrow B \bar B$, can help ferret them out.

\end{document}